\documentclass[aps,prl,twocolumn,groupedaddress,showpacs]{revtex4}
\usepackage{graphicx}
\draft
\begin{document}
\title{
Spin-Hall Effect in Two-Dimensional Electron Systems with Rashba
Spin-Orbit Coupling and Disorder }
\author{L. Sheng$^{1}$, D. N. Sheng$^2$ and C. S. Ting$^1$}
\address{
$^1$Department of Physics and Texas Center for Superconductivity,
University of Houston, Houston, Texas 77204\\
$^2$Department of Physics and Astronomy, California State
University, Northridge, California 91330 }

\begin{abstract}
Using the four-terminal Landauer-B\"{u}ttiker formula and Green's
function approach, we calculate numerically the spin-Hall
conductance in a two-dimensional junction system with the Rashba
spin-orbit (SO) coupling and disorder. We find that the spin-Hall
conductance can be much greater or smaller than the universal
value $e/8\pi$, depending on the magnitude of the SO coupling, the
electron Fermi energy and the disorder strength. The spin-Hall
conductance does not vanish with increasing sample size for a wide
range of disorder strength. Our numerical calculation reveals that
a nonzero SO coupling can induce electron delocalization for
disorder strength smaller than a critical value, and the
nonvanishing spin-Hall effect appears mainly in the metallic
regime.

\end{abstract}

\mbox{}\\

 \maketitle


The emerging field of spintronics,\cite{s1,s2} which is aimed at
exquisite control over the transport of electron spins in
solid-state systems, has attracted much recent interest. One
central issue in the field is how to effectively generate
spin-polarized currents in paramagnetic semiconductors. In the
past several years, many works~\cite{s1,s2,s3,s4,s5} have been
devoted to the study of injection of spin-polarized charge flows
into the nonmagnetic semiconductors from ferromagnetic metals.
Recent discovery of intrinsic spin-Hall effect in $p$-doped
semiconductors by Murakami $et$ $al.$~\cite{t1} and in Rashba
spin-orbit (SO) coupled two-dimensional electron system (2DES) by
Sinova $et$ $al.$~\cite{t2} may possibly lead to a new solution to
the issue. For the Rashba SO coupling model, the spin-Hall
conductivity is found to have a universal value $e/8\pi$ in a
clean bulk sample when the two Rashba bands are both occupied,
being insensitive to the SO coupling strength and electron Fermi
energy~\cite{t2}.

While the spin-Hall effect has generated much interest in the
research community,~\cite{t3,t6,tt7,t7,t9,t10,t12,t14,t15,t13}
theoretical works remain highly controversial regarding its fate
in the presence of disorder. Within a semiclassical treatment of
disorder scattering, Burkov $et$ $al.$~\cite{tt7} and Schliemann
and Loss~\cite{t7} showed that spin-Hall effect only survives at
weak disorder. On the other hand, Inoue $et$ $al.$~\cite{t12}
pointed out that the spin-Hall effect vanishes even for weak
disorder taking into account the vertex corrections. Mishchenko
$et$ $al.$~\cite{t14} further showed that the dc spin-Hall current
vanishes in an impure bulk sample, but may exist near the boundary
of a finite system. Nomura $et$ $al.$~\cite{t15} evaluated the
Kubo formula by calculating the single-particle eigenstates in
momentum space with finite momentum cutoff, and found that the
spin-Hall effect does not decrease with sample size at rather weak
disorder. Therefore, further investigations of disorder effect in
the SO coupled 2DES are highly desirable.

In this Letter, the spin-Hall conductance (SHC) in a 2DES junction
with the Rashba SO coupling is studied by using the four-terminal
Landauer-B\"{u}ttiker (LB) formula with the aid of the Green's
functions. We find that the SHC does not take the universal value,
and it depends critically on the magnitude of the SO coupling, the
electron Fermi energy, and the disorder strength. For a wide range
disorder strength, we show that the SHC does not decrease with
sample size and extrapolates to nonzero values in the limit of
large system. The numerical calculation of electron localization
length based upon the transfer matrix method also reveals that the
Rashba SO coupling can induce a metallic phase, and the spin-Hall
effect is mainly confined in the metallic regime. The origin of
the nonuniversal SHC in the 2DES junction is also discussed.

Let us consider a two-dimensional junction consisting of an impure
square sample of side $L$ connected with four ideal leads, as
illustrated in the inset of Fig.\ 1. The leads are connected to
four electron reservoirs at chemical potentials $\mu_0$, $\mu_1$,
$\mu_2$ and $\mu_3$. In the tight-binding representation, the
Hamiltonian for the system including the sample and the leads can
be written as~\cite{h1,Echo}
\begin{eqnarray}
H&=&-t\sum\limits_{\langle ij\rangle\sigma}c_{i,\sigma}^\dagger
c_{j,\sigma}+
\sum\limits_{i\sigma}\varepsilon_{i}c_{i\sigma}^\dagger
c_{i\sigma}\nonumber\\
&+&V_{\mbox{\tiny SO}}
\sum\limits_{i}\left[\left(c_{i,\uparrow}^\dagger
c_{i+\delta_x,\downarrow}-c_{i,\downarrow}^\dagger
c_{i+\delta_x,\uparrow}\right)\right.\nonumber\\
&-&\left.i\left(c_{i,\uparrow}^\dagger
c_{i+\delta_y,\downarrow}+c_{i,\downarrow}^\dagger
c_{i+\delta_y,\uparrow}\right)+\mbox{H.c.}\right]\ .\label{HAMIL}
\end{eqnarray}
Here, $V_{\mbox{\tiny SO}}$ is the SO coupling strength,
$\varepsilon_{i}\equiv 0$ in the leads and are uniformly
distributed between $[-W/2, W/2]$ in the sample, which accounts
for nonmagnetic disorder. The lattice constant is taken to be
unity, and $\delta_x$ and $\delta_y$ are unit vectors along the
$x$ and $y$ directions. In the vertical leads 2 and 3,
$V_{\mbox{\tiny SO}}$ is assumed to be zero in order to avoid
spin-flip effect, so that a probability-conserved spin current can
be detected in the leads.

The electrical current outgoing through lead $l$ can be calculated
from the LB formula~\cite{l1} $I_l=(e^2/h)\sum_{l'\neq
l}T_{l,l'}(U_{l'}-U_{l})$, where $U_{l}=\mu_l/(-e)$ and $T_{l,l'}$
is the total electron transmission coefficient from lead $l'$ to
lead $l$. A number of symmetry relations for the transmission
coefficients result from the time-reversal and inversion
invariance of the system after average of disorder configurations,
use of which will be implied. We consider that a current $I$ is
driven through leads $0$ and $1$, and adjust $U_l$'s to make
$I_1=-I_0=I$ and $I_3=I_2=0$. Since in the present system the
off-diagonal conductance $G_{xy}$ vanishes by symmetry,
$U_{0}-U_{1}$ equals to the longitudinal voltage drop caused by
the current flow $I$. In the vertical leads 2 and 3, where
$V_{\mbox{\tiny SO}}=0$, the electrical currents are separable for
the two spin subbands $I_{l}=I_{l\uparrow} + I_{l\downarrow}$ with
$\uparrow$ and $\downarrow$ for spins parallel and antiparallel to
the $z$-axis. The spin current is given by
$I^{(l)}_{sH}=[\hbar/2(-e)](I_{l\uparrow}-I_{l\downarrow})$. By
use of the LB formula, it is straightforward to obtain for the
transverse spin current
$I^{(3)}_{sH}=-I^{(2)}_{sH}=G_{sH}(U_{0}-U_{1})$. Here, the
proportional coefficient
\begin{equation}
G_{sH}=-\frac{e}{4\pi}(T_{3\uparrow,0}-T_{3\downarrow,0})\label{GSH}\
,
\end{equation}
is the SHC, where $T_{3\sigma,0}$ is the electron transmission
coefficient from lead 0 to spin-$\sigma$ subband in lead 3.
Equation\ (\ref{GSH}) can be calculated in terms of the
nonequilibrium Green's functions~\cite{l2,l3,l4}
$G_{sH}=-(e/4\pi)\mbox{Tr}(\Gamma_3\eta G^{r}\Gamma_0G^{a})$.
Here, $\eta=1$ and $-1$ in the spin-$\uparrow$ and
spin-$\downarrow$ subspaces, respectively, and
$\Gamma_{l}=i[\Sigma_{l}-(\Sigma_{l})^\dagger]$ with $\Sigma_{l}$
the retarded electron self-energy in the sample due to electron
hopping coupling with lead $l$. The retarded Green's function
$G^{r}$ is given by
\begin{equation}
G^{r}=\frac{1}{E-H_{\mbox{\tiny C}}
-\sum_{l=0}^{3}\left(\Sigma_{l}\right)}\ ,\label{GREENs}
\end{equation}
and $G^{a}=(G^{r})^\dagger$, where $E$ stands for the electron
Fermi energy, and $H_{\mbox{\tiny C}}$ is the single-particle
Hamiltonian of the central square sample only. The self-energies
can be first computed exactly by matching up boundary conditions
for the Green's function at the interfaces by using the transfer
matrices of the leads~\cite{l5}. The Green's function Eq.\
(\ref{GREENs}) is then obtained through matrix inversion. In our
calculations, $G_{sH}$ is always averaged over up to 5000 disorder
realizations, whenever $W\neq 0$.

In Fig.\ 1, the SHC $G_{sH}$ is plotted as a function of the
electron Fermi energy $E$ at fixed size $L=40$ for several
disorder strengths. The SHC is always an odd function of electron
Fermi energy $E$, and vanishes at the band center $E=0$. The
antisymmetric energy dependence of the SHC is similar to that of
the Hall conductance in a tight-binding model~\cite{dns1997}, and
originates from the particle-hole symmetry of the system. For
$E<0$ and $E>0$ the charge carriers are electron-like and
hole-like, respectively, and so make opposite contributions to the
SHC. With increasing $E$ from the band bottom $E\simeq -4t$,
except for a small oscillation due to the discrete energy levels
in the finite-size sample, $G_{sH}$ increases continuously until
$E$ is very close to the band center $E=0$. It is easy to see from
Fig.\ 1 that at weak disorder $W\lesssim t$ the calculated
$G_{sH}$ may be greater than the universal value, namely, $0.5$ in
our unit $e/4\pi$.
\begin{figure}
\includegraphics[width=2.0in]{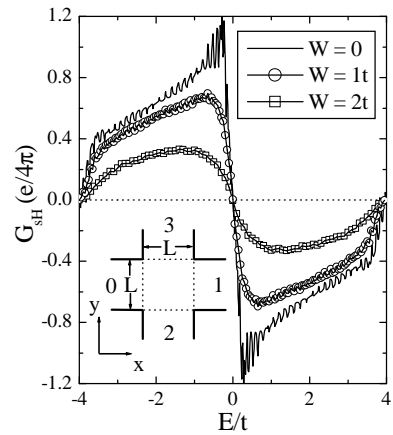}
\caption{Spin-Hall conductance $G_{sH}$ for some disorder
strengths as a function of electron Fermi energy $E$. Here, the
sample size $L=40$ and the spin-orbit coupling $V_{\mbox{\tiny
SO}}=0.5t$. Inset is a schematic view of the four-terminal
junction. }
\end{figure}

\begin{figure}
\includegraphics[width=2.0in]{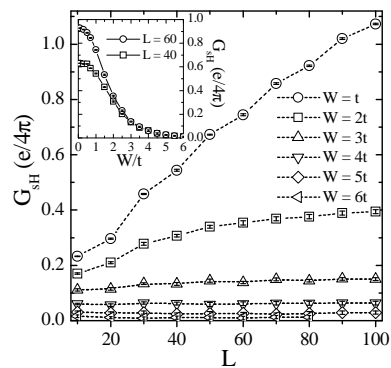}
\caption{Spin-Hall conductance as a function of sample size $L$
for different disorder strengths at $E=-2t$ and $V_{\mbox{\tiny
SO}}=0.5t$. Error bars due to statistical fluctuations, being
smaller than the symbol size, are drawn inside the open symbols.
Inset: spin-Hall conductance as a function of disorder strength
for $L=40$ and $60$. }
\end{figure}
In order to determine the behavior of the spin-Hall effect in
large systems, we calculate the SHC as a function of the sample
size from $L=10$ up to $100$ for different strengths of disorder,
as shown in Fig.\ 2. For weak disorder $W \lesssim 3t$, the SHC
first increases with increasing sample size, and then tends to
saturate. In particular, for $W\lesssim t$, we see that the SHC
can be several times greater than the universal value $e/8\pi$,
when the system becomes large. For a stronger disorder $3t
\lesssim W\lesssim 5t$, the SHC is roughly independent of the
sample size, and extrapolates to a finite value in the large-size
limit. Therefore, it is evident that the SHC will not vanish in
large systems in the presence of moderately strong disorder $W
\lesssim 5t$. With further increase of $W$, the SHC becomes
vanishingly small at $W \gtrsim 6t$, as seen more clearly from the
inset of Fig.\ 2, indicating that very strong disorder scattering
would eventually destroy the spin-Hall effect.

We further examine the dependence of the SHC on the strength of
the SO coupling. As shown in Fig.\ 3, overall, the SHC increases
with increasing $V_{\mbox{\tiny SO}}$ in the range $0\leq
V_{\mbox{\tiny SO}}\leq t$. For $W=0$ or weak disorder, the SHC
displays an interesting oscillation effect with a period much
greater than the average level spacing. According to Eq.\
(\ref{GSH}), the oscillation of the SHC is a manifestation of the
oscillation of the sideway spin-resolved transmission
coefficients. For a two-terminal junction with the SO coupling,
similar oscillation with finite sample size has previously been
observed for the spin-resolved transmission
coefficients~\cite{Echo}, where the oscillation period was
discussed to be the spin precession length $L_{sp}$. If we apply
the same condition $L=nL_{sp}$ with $n$ an integer and notice
$L_{sp}\simeq \pi t/V_{\mbox{\tiny SO}}$,~\cite{Echo} we can
obtain for the equivalent period in the SO coupling $\delta
V_{\mbox{\tiny SO}}\simeq \pi t/L$. For the parameters used in
Fig.\ 3, $\delta V_{\mbox{\tiny SO}}\simeq 0.08t$, which is very
close to the period as seen in the figure. This indicates that the
oscillation of the SHC is due to a spin precessional effect in
finite-size systems. Experimentally, $V_{\mbox{\tiny SO}}$ can be
varied over a wide range by tuning a gate voltage~\cite{r1,h3},
and so this oscillation effect may possibly be observed directly.
\begin{figure}
\includegraphics[width=2.0in]{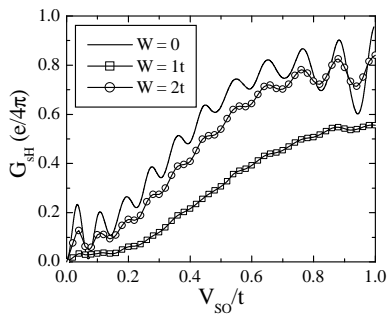}
\caption{Spin-Hall conductance as a function of spin-orbit
coupling strength for some disorder strengths. Here, the sample
size $L=40$ and the electron Fermi energy $E=-2t$. }
\end{figure}

Electron delocalization is a crucial issue for understanding
electron transport properties in the 2DES, and has already been
studied experimentally by use of magnetoresistance
measurements~\cite{h3}. For this reason, we investigate
numerically whether the Rashba SO coupling can induce a universal
electron delocalization in the presence of disorder. According to
the well-established transfer matrix approach,~\cite{h4,lisheng}
we calculate the electron localization length $\xi$ on a bar of
essentially infinite length ($5\times 10^5$) and finite width $L$.
In Fig.\ 4a, the normalized localization length $\xi/L$ is plotted
as a function of disorder strength for $V_{\mbox{\tiny SO}}=0.5t$
and $L=8$, 16, 32 and 64. At weak disorder, $\xi/L$ increases with
$L$, indicating that the localization length $\xi$ will diverge as
$L\rightarrow\infty$, corresponding to an electron delocalized
metallic phase. With the increase of $W$, $\xi/L$ goes down and
all the curves cross at a point (fixed point) $W=W_c\simeq 6.3t$,
where $\xi/L$ becomes independent of bar width $L$. For $W > W_c$,
$\xi/L$ decreases with $L$, indicating that $\xi$ will converge to
finite values as $L\rightarrow\infty$, corresponding to an
electron localized insulator phase. Thus the fixed point $W=W_c$
is the critical disorder strength for the metal-insulator
transition. Our result is consistent with the earlier calculation
by Ando~\cite{h1}, where a metallic phase was established at the
band center $E=0$ for a strong Rashba SO coupling.
\begin{figure}
\includegraphics[width=3.0in]{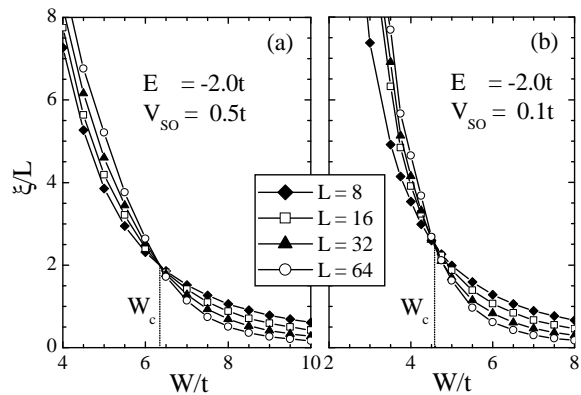}
\caption{ Normalized localization length as a function of disorder
strength calculated on long bars of length $5\times 10^5$ and
widths $L=8$, 16, 32 and 64. }
\end{figure}
Here, we also study weak SO coupling. In Fig.\ 4b, we plot the
result for a SO coupling strength much smaller than the electron
hopping integral, i.e., $V_{\mbox{\tiny SO}}=0.1t$, and similar
phase transition is also revealed at $W_c\simeq 4.6t$. In general,
we have performed calculations in the whole range from strong to
weak SO coupling (details will be presented elsewhere), and found
that electron delocalization occurs for any nonzero SO coupling
strength as the magnitude of the disorder varies.
Our result is in agreement with the perturbative
calculation of weak localization.~\cite{WeakLoc} As
$V_{\mbox{\tiny SO}}$ reduces, the critical $W_c$ decreases, and
the size-independent critical $\xi/L$ increases (so does the
critical longitudinal conductance $G_{xx}$~\cite{h4,lisheng}). In
the limit $V_{\mbox{\tiny SO}}\rightarrow 0$, we have
$W_c\rightarrow 0$ and all electron states become localized,
recovering the known regime of the two-dimensional Anderson model
for electron localization~\cite{h4}. The fact that the critical
$\xi/L$ changes with $V_{\mbox{\tiny SO}}$ indicates that the SO
coupled 2DES belongs to the universality class of two-parameter
scaling~\cite{NewDNS}. Comparing $W_{c}=6.3t$ calculated in Fig.\
4a for $V_{\mbox{\tiny SO}}=0.5t$ and $E=-2t$ with the SHC shown
in Fig.\ 2 for the same parameters, we see that nonvanishing
spin-Hall effect exists mainly in the metallic regime.

Our numerical study addresses the spin-Hall effect in a
finite-size junction system with leads. A comparison between the
spin-Hall effect and the quantum Hall effect (QHE) can shed some
light on the nonuniversal SHC obtained. For a QHE system,
delocalized states exist at the centers of the discrete Landau
levels, which are separated by mobility gaps consisting of
localized states. In the unit of conductance quantum $e^2/h$, the
Hall conductance is known to be a sum of the topological Chern
numbers of all the occupied delocalized states below the Fermi
energy~\cite{dns1997}. If the Fermi energy lies in a mobility gap,
the Hall conductance is well quantized to an integer. If the Fermi
energy is at a critical point, where a delocalized state exists,
the Hall conductance intrinsically fluctuates between two
integers. Similarly, the SHC is also related to corresponding
topological numbers of the occupied delocalized states. However,
in the present spin-Hall systems, the delocalized states
constitute a continuous spectrum without mobility gaps (or energy
gaps~\cite{h1}). Due to the lack of a mobility gap around the
Fermi energy, the SHC can fluctuate and does not show quantized
plateaus.  As a matter of fact, the universal value $e/8\pi$
predicted for clean bulk systems~\cite{t2} is 0.5 instead of an
integer in the unit of spin conductance quantum $e/4\pi$ (here the
electron charge $e$ in the conductance quantum $e^2/h$ needs be
replaced with electron spin $\hbar/2$). For the above reason, one
could not expect that the SHC will not change to different values
under different boundary conditions. In the present junction
system, the open boundary, i.e., the connection of the finite-size
sample with the much larger semi-infinite leads is quite different
from the essentially close boundary used in previous
calculations~\cite{t2,t6,tt7,t7,t9,t10,t12,t14,t15}, which is
likely the cause for the SHC to be possibly greater or smaller
than $e/8\pi$ depending on the electron Fermi energy, the disorder
strength and the magnitude of the SO coupling. Notably, the
analytical calculation~\cite{t14} also indicates that the contacts
between a sample and leads could enhance the generation of spin
currents. Our calculations provide an important evidence that the
proposed intrinsic spin-Hall effect~\cite{t1,t2} may be realized
experimentally in junction systems in the presence of disorder.

\textbf{Note added:} After initial submission of this paper, we
became aware of a couple preprints by Nikoli\'{c}, Z\^{a}rbo and
Souma and by Hankiewicz $et$ $al.$~\cite{nikoli}, where similar LB
formula calculations were carried out. Despite different parameter
values used, their results of nonuniversal SHC robust against
disorder scattering are consistent with ours.

\textbf{Acknowledgment:} The authors would like to thank A. H.
MacDonald, Q. Niu and J. Sinova for stimulating discussions. This
work is supported by ACS-PRF 41752-AC10, Research Corporation
Award CC5643, the NSF grant/DMR-0307170 (DNS), and also by a grant
from the Robert A. Welch Foundation (CST). DNS wishes to thank the
Aspen Center for Physics and Kavli Institute for Theoretical
Physics for hospitality and support (through PHY99-07949 from
KITP), where part of this work was done.

\end{document}